\documentclass{IEEEcsmag}
\pdfoutput=1
\usepackage{etex}

\usepackage[utf8]{inputenc}
\usepackage{paralist}
\usepackage[table]{xcolor}
\usepackage{booktabs} 
\usepackage{graphicx}
\usepackage[english]{babel}
\usepackage[autolanguage]{numprint}
\usepackage{wrapfig}
\usepackage{tabularx}
\usepackage{mdframed}
\usepackage{lipsum}
\usepackage{xspace}
\usepackage{cellspace}
\usepackage{balance}
\usepackage{hyperref}
\usepackage{multicol}
\usepackage{caption}
\usepackage{xcolor}
\usepackage{float}
\usepackage{color}
\usepackage{microtype}
\let\labelindent\relax
\usepackage[inline]{enumitem}
\usepackage{svg}
\usepackage{amssymb}

\newcommand{\revision}[1]{\textcolor{black}{#1}}

\jvol{XX}
\jnum{XX}
\paper{XX}
\jmonth{December}
\jname{IEEE Software}
\pubyear{2021}

\SetLipsumDefault{1}
\setlist[enumerate,1]{label=\textit{\alph*)}}

\newcommand{\TODO}[1]{\textcolor{orange}{todo: #1}}\newcommand\todo\TODO

\usepackage[frozencache]{minted}
\usepackage[many]{tcolorbox}
\tcbuselibrary{listings}
\tcbuselibrary{minted}
\usepackage{ifthen}
\usepackage{cleveref}

\crefname{customListing}{Listing}{Listings}

\newcommand{\listingsize}{\fontsize{7.5}{13.5}\selectfont}

\newtcbinputlisting[use counter=customListing, number format=\arabic]{\customListing}[4]{
  listing engine=minted,
  minted language=#1,
  listing file={#2},
  listing only,
  title={
    Listing \thetcbcounter\ifthenelse{\equal{#3}{}}{}: #3
  },
  minted options = {
        linenos,
        breaklines=false, 
        breakbefore=.,
        baselinestretch=0.8,
        fontsize=\listingsize,
        numbersep=2.8mm,
    },
    overlay={
        \begin{tcbclipinterior}
            \fill[gray!25] (frame.south west) rectangle ([xshift=4mm]frame.north west);
        \end{tcbclipinterior}
    },
  breakable,
  enhanced,
  fonttitle=\small,
  label=lst:#4
}

\lstdefinelanguage{diff}{
  language=java,
  basicstyle=\ttfamily\scriptsize,
  sensitive=true,
  numbers=none,
  morecomment=[f][\color{gray}][0]{diff},
  morecomment=[f][\color{gray}][0]{index},
  morecomment=[f][\color{blue}][0]{@@},
  morecomment=[f][\color{magenta}][0]{***},
  morecomment=[f][\color{violet}][0]{!},
  morecomment=[f][\color{red!60!black}][0]{-},
  morecomment=[f][\color{green!60!black}][0]{+},
  morecomment=[f][\color{magenta}][0]{---},
  morecomment=[f][\color{magenta}][0]{+++},
  morecomment=[f][\color{gray}][0]{Binary},
  morecomment=[f][\color{gray}][0]{Only},
  morecomment=[f][\color{gray}][0]{old},
  morecomment=[f][\color{gray}][0]{new},
  morecomment=[f][\color{gray}][0]{rename},
  morecomment=[f][\color{gray}][0]{similarity},
  morecomment=[f][\color{gray}][0]{deleted},
  morecomment=[f][\color{magenta}][0]{***************},
  morecomment=[f][\color{red!60!black}][0]<,
  morecomment=[f][\color{green!60!black}][0]>,
  morecomment=[f][\color{blue}][0]{0},
  morecomment=[f][\color{blue}][0]{1},
  morecomment=[f][\color{blue}][0]{2},
  morecomment=[f][\color{blue}][0]{3},
  morecomment=[f][\color{blue}][0]{4},
  morecomment=[f][\color{blue}][0]{5},
  morecomment=[f][\color{blue}][0]{6},
  morecomment=[f][\color{blue}][0]{7},
  morecomment=[f][\color{blue}][0]{8},
  morecomment=[f][\color{blue}][0]{9},
}[comments]

\newcommand{\approach}{{\sc R-Hero}\xspace}
\newcommand{\github}{{GitHub}\xspace}
\newcommand{\travisci}{{Travis CI}\xspace}

\begin{document}

\sptitle{Accepted for publication in IEEE Software}
\editor{Special Issue on Automatic Program Repair}

\title{A Software-Repair Robot based on Continual Learning}

\author{Benoit Baudry, Zimin Chen, Khashayar Etemadi, Han Fu, Davide Ginelli, Steve Kommrusch, Matias Martinez, Martin Monperrus, Javier Ron, He Ye, Zhongxing Yu}
\affil{}

\begin{abstract}
Software bugs are common and correcting them accounts for a significant part of costs in the software development and maintenance process. This calls for automatic techniques to deal with them. One promising direction towards this goal is gaining repair knowledge from historical bug fixing examples. Retrieving insights from software development history is particularly appealing with the constant progress of machine learning  paradigms and skyrocketing `big' bug fixing data generated through Continuous Integration (CI). In this paper, we present \approach, a novel software repair bot that applies continual learning to acquire bug fixing strategies from continuous streams of source code changes, implemented for the single development platform Github/Travis CI. We describe \approach, our novel system for learning how to fix bugs based on continual training, and we uncover initial successes as well as novel research challenges for the community. 
\end{abstract}

\maketitle

\chapterinitial{Introduction}

Developing software is a complex process that creates software which typically suffers from errors, such as null pointer exceptions and memory leaks. These errors can have severe consequences, ranging from customer dissatisfaction to the loss of human lives. Correcting the errors manually is notoriously tedious, difficult, and time-consuming, and automatically repairing them has long been a dream in programming.

Among the various proposals for automatic program repair, one particularly promising category leverages machine learning on big data. For instance, Prophet \cite{long2016automatic} learns a generic model of how natural a fix is, and  uses the model to rank patch candidates. Our goal in this article is to report on our experience in building the program repair bot \approach, which automatically performs machine learning on continuous integration builds.

Continuous Integration (CI) is deeply integrated into collaborative development platforms such as \github and Bitbucket. Typically, developers make changes to the code base (called commits) and a CI service compiles the code and runs the tests for each commit. This compilation and test execution phase is called a ``build''. Development using continuous integration results in a continuous  stream of builds.  For instance, in October 2018 alone, there were 9,495,908 \travisci builds created from 2,378,349 unique commits on \github \cite{durieux2019analysis}.
Our intuition is that these build streams provide unique insights on how skilled human developers fix bugs. As an example, \cref{lst:human-patch} shows a real \github commit that replaces the `null' object with `this' object and makes the CI build process successfully pass. This patch, drawn from the commit and build stream, is a piece of evidence that `null' is abused frequently, and that replacing it with the reference to the current instance is a viable patching strategy. Such a patch could be one data point for an ever learning software repair bot.

\customListing{diff}{training_point.diff}{A human patch from commit 528696fc of \href{https://github.com/BenjaminNitro/Project110/commit/528696fc8d57efec1141f641befef36d84006551}{BenjaminNitro/Project110}.}{human-patch}

\begin{figure*}
    \includegraphics[width=2\columnwidth]{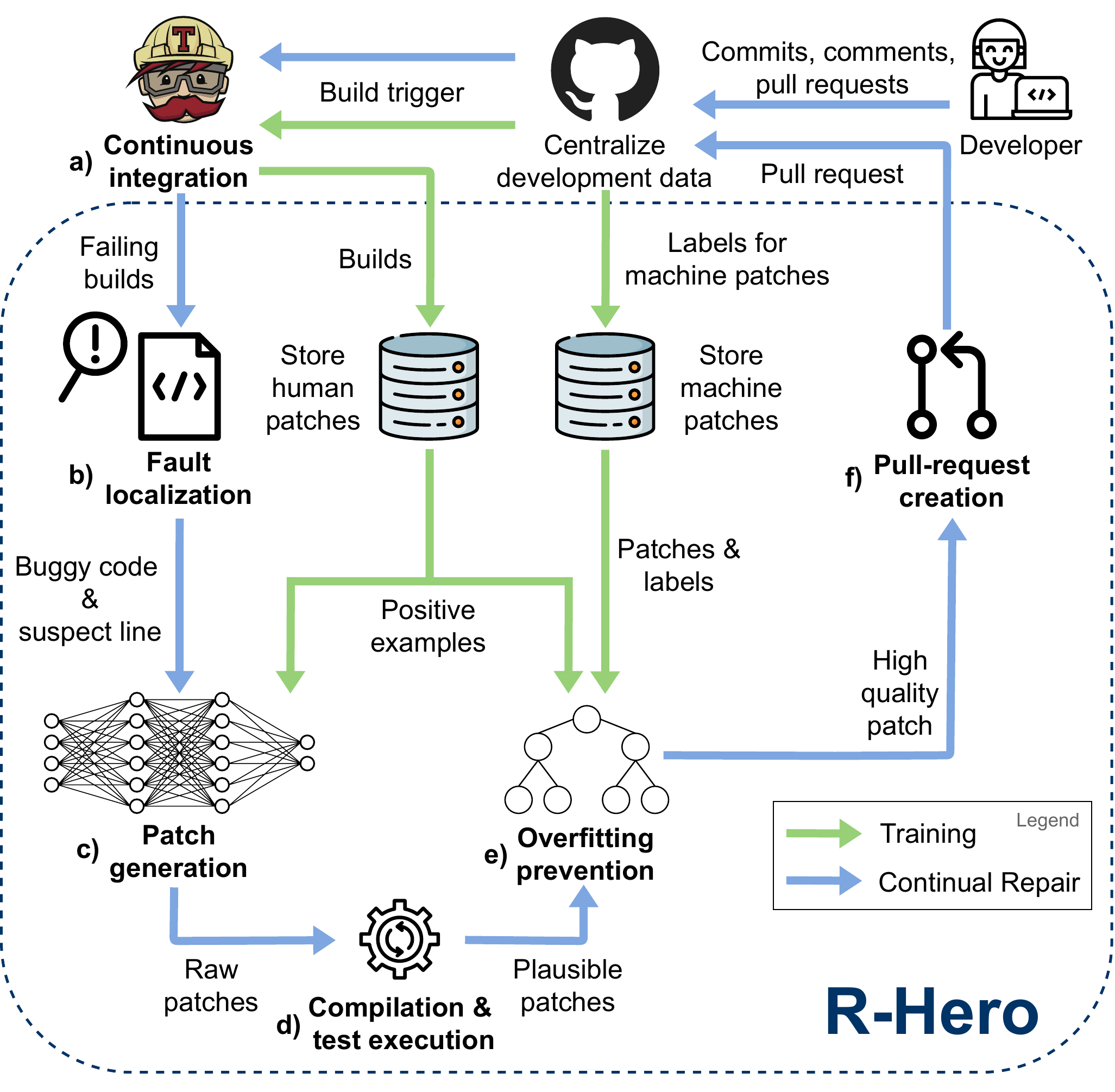}
    \captionsetup{width=2\columnwidth, margin={0\columnwidth, 0\columnwidth}, justification=centering}
    \captionof{figure}{Overview of \approach, a software bot that learns to generate patches, build after build.}
    \label{fig:overview}
\end{figure*}

Continual learning (also known as lifelong learning) is the learning paradigm which consists of learning continuously and adaptively from a stream of data.
Continual learning is viewed as a fundamental step towards artificial intelligence~\cite{parisi2019continual}, and research into this learning paradigm has produced groundbreaking results in the past few years. For example, in natural language processing, \revision{chatbots using continual learning have solved the underlying open-world knowledge problem \cite{chen2018lifelong}}, which consists of learning in conversations with facts that have never been encountered before.

Our key intuition is that, by feeding continuous integration build streams to continual learning techniques, the resultant model can potentially have expertise for fixing different types of bugs in an open manner. 
In the past years, we have worked on this problem and designed the first ever program repair bot based on continual learning. We call this bot \approach. 
\revision{\approach is designed to repair a specific class of errors: those that can be fixed with a single-line code change (aka one-liner). This class of errors has the following desirable properties:
it is a common class of errors, as witnessed by the fact that one-line commits constitute a significant portion of all commits \cite{ArafatR09};
these errors cover a wide range of behavioral problems such as control flow errors (fixing conditions), exception errors, missing behavior (addition of one single method call), etc \cite{zhong2015empirical};
one-line fixing is on the frontier of program repair research, as the large majority of systems focus on such bugs \cite{le2012genprog,nguyen2013semfix,long2016automatic,bader2019getafix,sequencer,SapFix2019}.}
In the following, we present the architecture for \approach, we report on our recent results on fixing real-life bugs in continuous integration, and analyze the research challenges uncovered by our experiments.

\subsection{Software Repair Bots}

Program repair techniques differ in various ways, including the type of oracle they use, as well as the targeted bug category \cite{LeGouesCACM}. Several techniques use test suites as the oracle and are designated as \emph{test suite based repair}.
These repair techniques are   classified into three categories: heuristic-based~\cite{le2012genprog}, synthesis-based~\cite{nguyen2013semfix}, and learning-based~\cite{long2016automatic}. They differ in how they analyze and modify the buggy code. \revision{ Due to both the advancements in machine learning and the massive amount of available CI data, learning based techniques are increasingly popular. Techniques in this category have been developed, for instance, to fix compilation errors \cite{deepfix} and learn meaningful patch changes \cite{codechange}.} 
One of the hardest problems of test suite based repair is that test suites are incomplete.
Consequently, the generated fix can pass the available tests but break untested functionality. This problem is called the overfitting issue \cite{qi2015efficient}. Current approaches typically use test augmentation to alleviate this problem. 

The integration of core repair algorithms into modern development workflows is a key dimension of the program repair problem space.
One possible integration is through software bots:
SapFix \cite{SapFix2019} and Repairnator~\cite{urli2018design} are examples of bots that constantly monitor software failures and run program repair tools against each bug.
Repairnator has produced human-competitive patches, which were accepted by human developers and permanently merged into the code base \cite{monperrus2019repairnator}. \approach represents our newest effort in pushing the state of the art of program repair bots: compared to Repairnator, \approach makes full use of ``Big Commit Data'' based on continual learning to ever improving its repair capabilities.

\subsection{Continual Learning for Repair}

We deploy continual learning on top of continuous integration build data streams.
We argue that continual learning is an appropriate paradigm for a program repair bot because code  constantly evolves. New libraries and tools get developed, new approaches to programming problems emerge, and new security patching strategies are uncovered.
By using continual learning, the idea is that the overall performance of the repair bot improves over time, because the repair model is never set in stone.
Continual learning would be able to capture bug fixing strategies that occur over different time frames, from days (emergency patching of 0-day vulnerabilities) to months (API updates) and years (programming language evolution).

\subsection{Architecture of \approach}

Figure~\ref{fig:overview} shows the six main building blocks of \approach:
\begin{inparaenum}[\it a)]
\item {Continuous integration},
\item {Fault localization},
\item {Patch generation},
\item {Compilation \& Test execution},
\item {Overfitting prevention},
and
\item {Pull-request creation}.
\end{inparaenum}
\approach stores its knowledge in two databases respectively composed of human-written and machine-synthesized patches.

\approach receives and analyzes the events from a continuous integration (CI) system  such as \travisci.
It collects commits that result in a passing build as determined by CI.
The changes from a commit may or may not have been a bug fix, but the fact that the change passes all tests hints that it is useful training data.
\revision{The extraction of single-line changes works as follows: for each commit, \approach goes over the corresponding \textit{diff} and iterates over each \textit{hunk}. It extracts training data only from hunks that describe single-line changes. This may produce several useful training data points per commit.}
\revision{In other words, \approach uses the before and after commit code to train its machine learning model for patch generation}.
\approach currently relies on the SequenceR ML-based patch generator~\cite{sequencer}, a sequence-to-sequence neural network model trained to receive buggy code as input and to generate patch proposals as output.
At each training step, SequenceR updates its model's weights to determine the tokens that should be output in the proposed patches.

Next, we detail the repair process, shown with blue arrows in Figure~\ref{fig:overview}. The repair process is triggered by 
\approach monitoring the continuous integration to detect failing builds, manifested by at least one failing test case.
For a given failing build, \approach checks out the version of the project that produces the failing build. 
Then, the fault localization component models the program under repair and pinpoints the locations that could be buggy (file names and line numbers).
\approach passes the collected locations to SequenceR which generates one or more potential patches for each location. 
Because the patch generation was trained on any kind of one-line change that results in a passing build, \approach repairs both compilation errors and test failures. 

\approach then validates each candidate patch.
It first compiles the patches and  executes all the tests to verify if, after applying the patch, the compilation and test execution do not fail anymore.
The patches that pass both validations are known as \emph{plausible} patches.
Once \approach finds \emph{plausible} patches, it assesses them, in order to avoid annoying developers with \emph{overfitting patches}.
This check is based on the overfitting detection system \emph{ODS} \cite{odsArxiv}.
ODS is a probabilistic model trained using supervised learning on both human patches (which are assumed to be positive examples), and machine patches (labelled as correct or incorrect), collected from previous program repair research~\cite{ye2019RGT}.

Finally, when a high-quality patch is identified by ODS, \approach submits a pull-request to the corresponding \github project which has the failing build.  The pull-request message to the developer describes the build failure and the patch, \href{https://bit.ly/3fRIHhd}{https://bit.ly/3fRIHhd} is an example of such a pull-request.

\section{Achievements of \approach}
\label{sec:data_collection}

\subsection{Data Collection}
In order to measure the applicability of \approach, we started it from scratch on May 16, 2020. At this date, all weights in the neural network were random, and \approach did not know anything about how to generate patches.
Then, we started to analyze the stream of continuous integration build data: looking at passing builds to train the system in a continual manner, and trying to generate patches for failing builds. \revision{For this purpose, \approach constantly monitors all \travisci and \github Actions builds for \github repositories that use Maven. Consequently, \approach does not operate on a fixed set of projects, it can potentially repair any failed build in any Github project.}
We continued from this date until \approach had synthesized plausible patches for projects from 10 different \github organizations (in \github, an \emph{organization} is a shared account for groups and companies) and until a fully automated pull-request had been created. Overall, this execution took $196$ days and ended on Nov 28, 2020. At this point \approach had collected 550,000 one-line code changes for training and, based on the trained model, \approach had tried to repair 44,002 failing builds that were reproduced locally.

We complement this automated patch collection with manual analysis. To check whether \approach generates human-like fixes for build failures, we manually collect the fixes implemented by developers. 
We find potential developer fixes as follows: the first commit with a successful build happening after the failed one. If the manual analysis of this last pushed commit shows that it does resolve the failure, we consider it as the ground truth fix made by human developers. 
This manual analysis revealed that one patch synthesized by \approach is identical to the patch produced by the human developer, described in the Patch Story section next.

\subsection{Story of a correct patch by \approach}

\emph{Lizzie} is an open-source project hosted on \github. It has 22k lines of code written by more than 30 developers along 1,030 commits, and has been praised with more than 600 \github stars.
On the morning of Aug 19, 2020, one commit to \emph{Lizzie} broke the \travisci build.
That commit, which modifies the file \texttt{GIBParser.java}, moved the declaration of the variable \texttt{sk} to below its first use.
That change produced a compilation error: \emph{``cannot find symbol: variable sk''}. As a result, \travisci could not build the revision.

\customListing{diff}{patch.diff}{\approach Patch for build 719254693, identical to the developer patch, generated after training over 70,914 commits.}{r-hero_patch}

The same morning, \approach watched and analyzed the status of \emph{thousands} of \travisci's builds, including that one from \emph{Lizzie}. 
As soon as \approach detected the broken status of that build, it started its repair mission.
\approach first checked out the failing project and locally reproduced the failure.
Then, \approach passed the \emph{Lizzie} project to the repair pipeline to find candidate patches. 
At this point in time, \approach continual learning model had been trained with data extracted from 70,914 commits.
\approach's patch generation output one patch, displayed in \cref{lst:r-hero_patch}: it replaces the variable \texttt{`sk'} (not yet declared at line 6) by the variable \texttt{`i'}, declared a few lines above.
\approach verified that the patched program compiled and passed all \emph{Lizzie}'s unit tests.
In total, \approach took 8 minutes to execute all of the mentioned steps.

Two days later, Aug 21 2020, a developer committed a patch that fixed the problem introduced on Aug 19.
The patch is identical to the one automatically synthesized by \approach.
\approach learned from zero knowledge to create that patch by only observing the build stream.
It is worth noting that \emph{Lizzie}, the project repaired by \approach, is the UI of \emph{Leela Zero}, an open-source implementation of AlphaGo Zero, a famous system that learns, also from zero, to play Go.

\subsection{Overall Performance}

\emph{Patch Diversity.}
\approach was able to synthesize 85 plausible patches to fix 13 failed builds and made one pull-request. Four of these 13 builds failed due to a compilation error, and nine failed due to test failures. This demonstrates that \approach is indeed able to fix both types of bugs (compilation errors and test failures). We systematically analyzed the 85 patches created by \approach and found that it was able to produces six types of patches:
\begin{enumerate*}
    \item return expression update~(21),
    \item method invocation update~(25),
    \item assignment update~(12),
    \item if condition update~(26), and
    \item removing a \textit{try} keyword~(1).
\end{enumerate*}
This diversity of bug fixing strategies confirms that data-driven repair is effective \cite{bader2019getafix}, and shows that \approach captures the variety of problems behind continuous integration build failures.

\begin{figure}[H]
{
    \vspace{0.2cm}
    \centering
    \includesvg[width=\columnwidth]{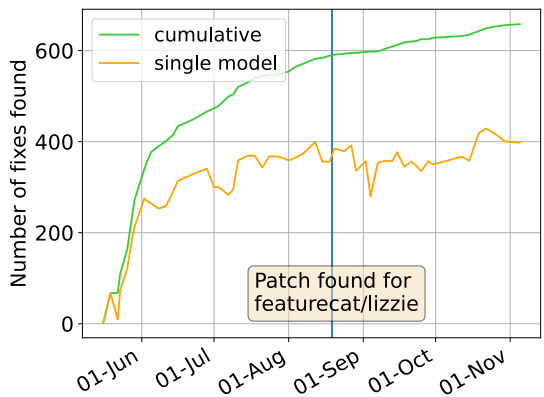}
    \captionsetup{justification=centering}
    \captionof{figure}{\revision{The orange line shows the number of CodRep4 fixes correctly produced by a single model trained on the data available to that point in time~\cite{sequencer}. The green line shows the cumulative number of fixes aggregated over all models. The decrease that regularly happens shows that a single model sometimes forgets how to fix bugs previously fixed.}}
    \label{fig:repair}
    \vspace{0.2cm}
}
\end{figure}

\emph{Performance over time.}
We conduct an experiment based on the protocol of Chen et al. \cite{sequencer} to measure the overall performance of \approach. The patch dataset, called CodRep4, consists of 3998 pairs of \emph{buggy} and \emph{fixed} versions of Java source code. Each time the \approach model is updated with continual learning (i.e. every three days approximately), we count the number of bugs for which \approach generates a patch that is syntactically identical to the corresponding fixed version. Note that due to continual learning, the \approach model changes over time, therefore, \approach produces different numbers of correct patches at different points in time.
The results of this experiment are shown in \autoref{fig:repair}. Each point on the orange curve indicates the number of correct CodRep4 patches that \approach generates at a specific point in time. Each point on the green line shows the cumulative number of correct patches generated by \approach. For example, the execution on July 1st resulted in correct patches for 300 buggy files in CodRep4. 
The increasing trends in \autoref{fig:repair} confirm that continual learning is a valid approach for program repair.

\emph{Overfitting Patch Classification.}
Currently, \approach's overfitting detection module uses our best-in-breed model trained from a) \numprint{2003} human-written correct patches and b) \numprint{8299} automatically synthesized incorrect patches, labelled by humans and the automated technique RGT~\cite{ye2019RGT}.
Of the 85 plausible patches, \approach discarded 64 incorrect patches, showing that it is able to increase the precision of generated patches reported to the developer. To further study this, we have performed an in-depth manual analysis for all of 64 patches. Among 64 patches, we identified that 57 patches are overfitting, which means ODS’ classification was correct in 89.06\% of cases on this data, confirming previous research \cite{odsArxiv}.
Overall, ODS and SequenceR work well in concert n R-Hero, discarding the majority of the overfitting patches and increasing the value proposition for the developer.

\subsection{End-to-end Integration}

\approach integrates many different state-of-the-art components from program repair research, and the question of the feasibility of fully automated pull-requests is an open one in the community~\cite{urli2018design}. 
\approach demonstrates that this is the case.
On Nov 28, 2020, after 196 days of training, 
\approach created a first pull-request to the \emph{thomasleplus/xee} project (\url{https://github.com/thomasleplus/xee}) without any human intervention:
the data collection for training SequenceR, the patch synthesis, the validation of the patch using ODS and the mechanism to propose the patch to the developer were done by \approach in a fully automated manner.
This pull-request proves that \approach is able to learn to repair a failing build from zero, starting with zero repair knowledge.
The proposed pull-request was not merged by the developer, stating that it does not fully fix the bug, yet, according to the developer feedback, the patch exposed incorrect design in the failing test case.

\subsection{Reproducibility}

All the data of this experiment has been systematically saved for sake of scientific reproducibility, it is available on a \github repository (\href{https://github.com/repairnator/open-science-repairnator/}{github.com/repairnator/open-science-repairnator/}). For every build, we provide the \travisci data, the log file associated with the execution of \approach, and the plausible patches created by \approach. For every patch, we provide its manual analysis categorization about the failure and correctness data. 

\section{Research Challenges}
In this section, we summarize the two important research challenges highlighted by \approach. 

\subsection{Overfitting Patches and Compilation Errors}

\revision{
To our knowledge, most of the research about overfitting research focuses on test-based program repair. Consequently, ``overfitting patches'' mostly refer to patches that pass developer-provided test cases, but don't correctly fix the bug. Our work reveals that overfitting can also affect compilation bugs: we have observed that \approach may find a patch  that perfectly repairs a compilation error while being incorrect. This shows that considering the compiler as an oracle is flawed, contrary to the core assumption of the related research on compilation error repair \cite{deepfix}. In short, the research challenge is: what kind of additional oracles can be used to avoid degenerated patches in compilation error repair?}  

\subsection{Catastrophic Forgetting}

Continual learning is prone to a problem called catastrophic forgetting, which occurs when newly learned knowledge interferes with capabilities previously learned by the model \cite{parisi2019continual}. This phenomenon causes the model to forget old knowledge and potentially leads to decreased performance. In \autoref{fig:repair}, the orange line represents the performance of \approach evaluated on a reference dataset. As we can see, during the initial stage, the performance steadily improves, but \approach stops improving after one month of continual learning. Then, the performance varies and even sometimes decreases. \revision{However, the green line in \autoref{fig:repair} (cumulative number of correct fixes) shows that \approach is still learning to repair new bugs. This indicates that the non-monotonic increase of the the orange line is due to \approach forgetting previously learned bug fixes, i.e., catastrophic forgetting is happening. Our experiment is the first one to show that catastrophic forgetting happens in the context of learning-based program repair, and our initial investigation of the problem reveals that it is a hard research problem.}

\section{Limitations}
\revision{Though R-Hero shows great promise in fixing software bugs happening in continuous integration, its limitations clearly call for future work: 1) R-Hero is  trained on one-line changes, this limits  R-Hero from fixing complex bugs requiring multi-line edits (as for most existing repair systems). 2) the prototype implementation of R-Hero does not scale: among the considered 44,002 failing builds, R-Hero has only repaired 13/44,002 builds.  This is arguably a low ratio but, to the best of our knowledge, nobody has ever succeeded in reporting a higher ratio on arbitrary builds from continuous integration (i.e. by drawing from the field distribution of all possible bugs). So far, R-Hero has done a single pull-request, discussed above, and we hope it is the beginning of a fruitful series of bot contributions.}

\section{Conclusion}

We believe that continual learning and continuous feedback are essential ingredients to go beyond pure software technology. They provide a useful paradigm to incorporate human knowledge into a self-improving software system, and for automated repair, they enable software engineering research to achieve a truly socio-technical repair system. To that extent, \approach is a milestone in showing that developers and bots can cooperate fruitfully to produce high-quality, reliable software systems.

\textbf{Acknowledgments} This work was partially supported by the Wallenberg Artificial Intelligence, Autonomous Systems and Software Program (WASP) funded by Knut and Alice Wallenberg Foundation, by the Swedish Foundation for Strategic Research (SSF). The experiments were performed on resources provided by the Swedish National Infrastructure for Computing.

\bibliographystyle{plain}
\bibliography{bibliography} 

\begin{thebibliography}{10}

\bibitem{ArafatR09}
Oliver Arafat and Dirk Riehle.
\newblock The commit size distribution of open source software.
\newblock In {\em 42st Hawaii International International Conference on Systems
  Science {(HICSS-42} 2009), Proceedings {(CD-ROM} and online), 5-8 January
  2009, Waikoloa, Big Island, HI, {USA}}, pages 1--8. {IEEE} Computer Society,
  2009.

\bibitem{bader2019getafix}
Johannes Bader, Andrew Scott, Michael Pradel, and Satish Chandra.
\newblock Getafix: Learning to fix bugs automatically.
\newblock {\em Proceedings of the ACM on Programming Languages},
  3(OOPSLA):1--27, 2019.

\bibitem{chen2018lifelong}
Zhiyuan Chen and Bing Liu.
\newblock Lifelong machine learning.
\newblock {\em Synthesis Lectures on Artificial Intelligence and Machine
  Learning}, 12(3):1--207, 2018.

\bibitem{sequencer}
Zimin Chen, Steve Kommrusch, Michele Tufano, Louis-Noël Pouchet, Denys
  Poshyvanyk, and Martin Monperrus.
\newblock Sequencer: Sequence-to-sequence learning for end-to-end program
  repair.
\newblock {\em IEEE Transactions on Software Engineering}, 2019.

\bibitem{durieux2019analysis}
Thomas Durieux, Rui Abreu, Martin Monperrus, Tegawend{\'e}~F Bissyand{\'e}, and
  Lu{\'\i}s Cruz.
\newblock An analysis of 35+ million jobs of travis ci.
\newblock In {\em 2019 IEEE International Conference on Software Maintenance
  and Evolution (ICSME)}, pages 291--295. IEEE, 2019.

\bibitem{LeGouesCACM}
Claire~Le Goues, Michael Pradel, and Abhik Roychoudhury.
\newblock Automated program repair.
\newblock {\em Communications of the ACM}, 2019.

\bibitem{deepfix}
Rahul Gupta, Soham Pal, Aditya Kanade, and Shirish Shevade.
\newblock Deepfix: Fixing common c language errors by deep learning.
\newblock In {\em Proceedings of the Thirty-First AAAI Conference on Artificial
  Intelligence}, AAAI'17, page 1345–1351. AAAI Press, 2017.

\bibitem{le2012genprog}
Claire Le~Goues, ThanhVu Nguyen, Stephanie Forrest, and Westley Weimer.
\newblock Genprog: A generic method for automatic software repair.
\newblock {\em Ieee transactions on software engineering}, 38(1):54--72, 2012.

\bibitem{long2016automatic}
Fan Long and Martin Rinard.
\newblock Automatic patch generation by learning correct code.
\newblock In {\em Proceedings of the 43rd Annual ACM SIGPLAN-SIGACT Symposium
  on Principles of Programming Languages}, pages 298--312, 2016.

\bibitem{SapFix2019}
Alexandru Marginean, Johannes Bader, Satish Chandra, Yue~Jia Mark~Harman,
  Ke~Mao, Alexander Mols, and Andrew Scott.
\newblock Sapfix: Automated end-to-end repair at scale.
\newblock In {\em International Conference on Software Engineering}, 2019.

\bibitem{monperrus2019repairnator}
Martin Monperrus, Simon Urli, Thomas Durieux, Matias Martinez, Benoit Baudry,
  and Lionel Seinturier.
\newblock Repairnator patches programs automatically.
\newblock {\em Ubiquity}, 2019(July):1--12, 2019.

\bibitem{nguyen2013semfix}
Hoang Duong~Thien Nguyen, Dawei Qi, Abhik Roychoudhury, and Satish Chandra.
\newblock Semfix: Program repair via semantic analysis.
\newblock In {\em 2013 35th International Conference on Software Engineering
  (ICSE)}, pages 772--781. IEEE, 2013.

\bibitem{parisi2019continual}
German~I Parisi, Ronald Kemker, Jose~L Part, Christopher Kanan, and Stefan
  Wermter.
\newblock Continual lifelong learning with neural networks: A review.
\newblock {\em Neural Networks}, 113:54--71, 2019.

\bibitem{qi2015efficient}
Zichao Qi, Fan Long, Sara Achour, and Martin Rinard.
\newblock An analysis of patch plausibility and correctness for
  generate-and-validate patch generation systems.
\newblock In {\em Proceedings of ISSTA}. ACM, 2015.

\bibitem{codechange}
Michele Tufano, Jevgenija Pantiuchina, Cody Watson, Gabriele Bavota, and Denys
  Poshyvanyk.
\newblock On learning meaningful code changes via neural machine translation.
\newblock In {\em Proceedings of the 41st International Conference on Software
  Engineering}, ICSE '19, page 25–36. IEEE Press, 2019.

\bibitem{urli2018design}
Simon Urli, Zhongxing Yu, Lionel Seinturier, and Martin Monperrus.
\newblock How to design a program repair bot? insights from the repairnator
  project.
\newblock In {\em 2018 IEEE/ACM 40th International Conference on Software
  Engineering: Software Engineering in Practice Track (ICSE-SEIP)}, pages
  95--104. IEEE, 2018.

\bibitem{odsArxiv}
He~Ye, Jian Gu, Matias Martinez, Thomas Durieux, and Martin Monperrus.
\newblock Automated classification of overfitting patches with statically
  extracted code features.
\newblock {\em IEEE Transactions on Software Engineering}, (under press), 2021.

\bibitem{ye2019RGT}
He~Ye, Matias Martinez, and Martin Monperrus.
\newblock Automated patch assessment for program repair at scale.
\newblock {\em Empirical Software Engineering}, 1 2020.

\bibitem{zhong2015empirical}
Hao Zhong and Zhendong Su.
\newblock An empirical study on real bug fixes.
\newblock In {\em 37th IEEE International Conference on Software Engineering},
  volume~1, pages 913--923. IEEE, 2015.

\end{thebibliography}
\balance

\end{document}